\documentclass{ws-procs9x6}
\usepackage{amssymb,epsfig}

\begin{document}

\title{Neutron background at Boulby mine}

\vspace{-1.0cm}

\author{V.~A. KUDRYAVTSEV, P.~K. LIGHTFOOT, J.~E. McMILLAN, \\
M. ROBINSON and N.~J.~C. SPOONER}

\address{Department of Physics and Astronomy, University of Sheffield,
Sheffield, \\
S3 7RH, UK}

\vspace{-0.3cm}
\author{P.~F. SMITH, N.~J.~T. SMITH and J.~D. LEWIN}

\address{Particle Physics Department, 
Rutherford Appleton Laboratory, Chilton, Oxon \\
OX11 0QX, UK}  

\vspace{-0.3cm}
\author{R. L\"USCHER and I. LIUBARSKY}

\address{Blackett Laboratory, Imperial College of Science, 
Technology and Medicine, London SW7 2BZ, UK}


\maketitle

\vspace{-0.5cm}
\abstracts{
The neutron background at the underground laboratory at Boulby - a 
site for several dark matter experiments - is
discussed. Special emphasis is put on the neutron background
produced by cosmic-ray muons. The most recent versions of the muon 
propagation 
code MUSIC, and particle transport code FLUKA are used to 
evaluate muon and neutron fluxes. The results of simulations are
compared with experimental data.}

\vspace{-0.5cm}
\section{Introduction}
Neutrons are the major background in experiments 
searching for WIMP dark matter particles -- also known as neutralinos.
WIMPs are expected to interact with ordinary 
matter in detectors to produce nuclear recoils, which can be 
detected through ionisation, scintillation or phonons. Identical 
events can be induced by neutrons. Thus, only suppression of any background
neutron 
flux by passive or active shielding will allow experiments to 
reach sufficiently high sensitivity to neutralinos. Designing shielding 
for such 
detectors requires simulation of neutron fluxes.

Neutrons underground arise from two sources: i) local 
radioactivity, and ii) cosmic-ray muons. Neutrons associated with 
local radioactivity are produced mainly via ($\alpha$, n) reactions, 
initiated by $\alpha$-particles from U/Th traces in the rock and 
detector elements. Neutrons from spontaneous fission of $^{238}$U 
contribute also to the flux at low energies. The neutron yield from 
cosmic-ray muons depends strongly on the depth of the underground 
laboratory. At Boulby it is about 3 orders of magnitude less 
than that of neutrons arising from local radioactivity. 
A larger thickness of rock suppresses the muon flux and, 
hence, reduces also the neutron yield. The dependence, however, 
is not linear. The muon-induced neutron flux can
be important for experiments intending to reach high 
sensitivity to WIMPs or to low-energy neutrino fluxes. There are 
several reasons for this: 1) the energy spectrum of muon-induced neutrons 
is hard, extending to GeV energies, and fast neutrons can travel far from 
the associated
muon track, reaching a detector from large distances; 2) fast neutrons 
transfer larger energies to nuclear recoils making them visible in 
dark matter detectors, while many recoils from $\alpha$-induced neutrons
fall below detector energy thresholds; 3) a detector 
can be protected against neutrons from the rock activity by 
hydrogen-rich material, possibly with addition of thermal neutron 
absorber; such a material, however, will be a target for cosmic-ray 
muons and will not protect against muon-induced neutrons. The only way 
to reduce this flux is to add an active muon and/or neutron veto
(see also Ref. \cite{alex} for discussion).

\vspace{-0.3cm}
\section{Neutrons from the radioactivity in Boulby rock}
\begin{figure}[htb]
\mbox{
\epsfig{file=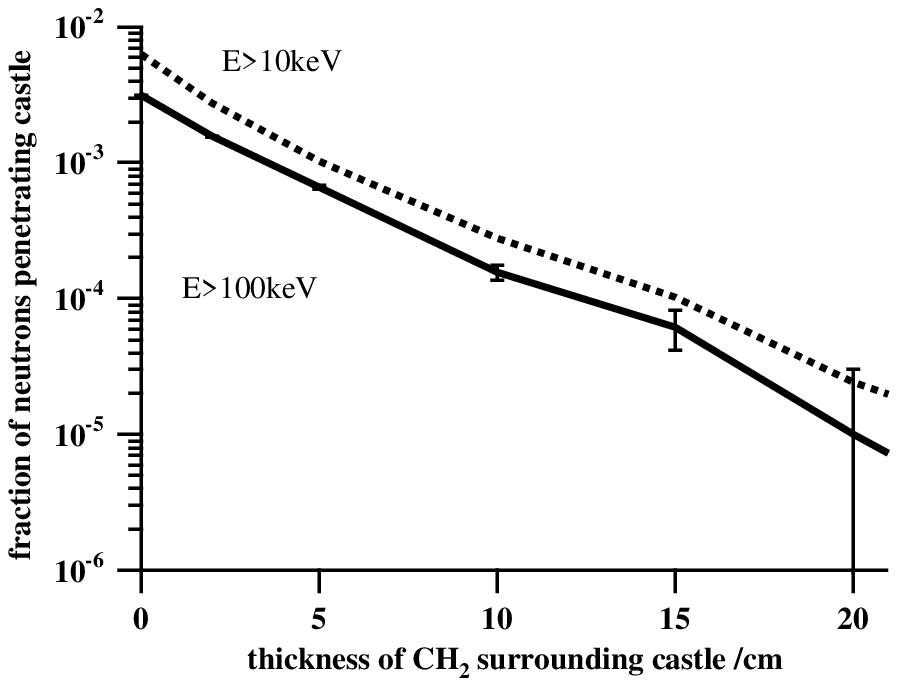,width=5.2cm,height=4.0cm}
\hspace{0.5cm}
\epsfig{file=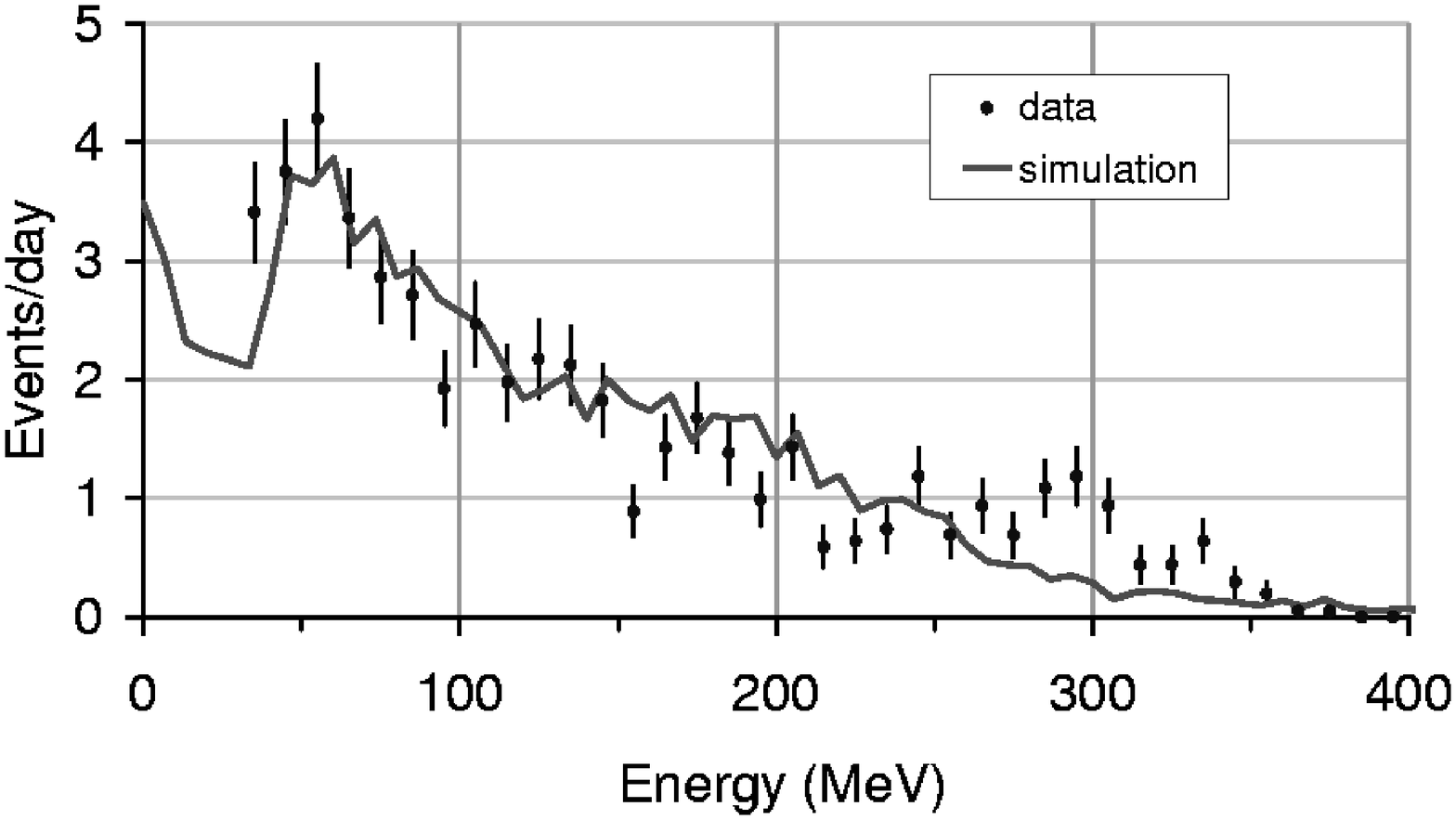,width=5.2cm,height=4.0cm}
}
\parbox{5.2cm}{\vspace{-0.0cm}\caption
{Fraction of survived neutrons above 10 keV and 100 keV as a function of
CH$_{2}$ thickness (behind 15 cm of lead and 10 cm of copper).
\label{fig:rockneutrons}}}
\hspace{0.7cm} \parbox{5.2cm}{\vspace{-0.0cm}\caption
{Muon energy deposition spectrum: measurements (filled 
circles and simulations (solid curve) \label{fig:muonspectrum}}}
\end{figure}

Several samples of rock in the Boulby mine around the experimental 
labs (mainly halite) were used to measure the contamination of U and 
Th. It was found that the fractions of these elements do not exceed 
60 ppb of U and 300 ppb of Th, but show large variations from 
sample to sample, being on average about 30 ppb of U and 150 ppb of 
Th. This gives about $2 \times 10^{-8}$ neutrons/g/s in halite. 
Neutron propagation with MCNP \cite{mcnp} results in a preliminary 
estimate of the
neutron flux behind the lead (15 cm) and copper (10 cm) shielding of 
about $10^{-6}$ cm$^{-2}$~s$^{-1}$ above 100 keV. 
Figure \ref{fig:rockneutrons} shows 
the absorption of the flux as a function of the thickness of the 
CH$_{2}$ material, commonly used to shield detectors from 
neutrons. 10 cm of CH$_{2}$ suppresses the neutron flux from the rock 
by about one order of magnitude.

\vspace{-0.3cm}
\section{Muon flux and energy spectrum underground}
Knowledge of the muon flux is clearly 
important for calculations of the muon-induced neutron flux or 
counting rate -- it provides the absolute normalisation.
The muon energy spectrum at a given experimental site also affects the 
neutron production rate. Thus precise 
knowledge of the muon spectrum and absolute normalisation is crucial for 
neutron flux simulations.

The measurement of the muon flux at Boulby was carried out using the 
liquid scintillator veto system of the ZEPLIN I dark matter detector 
\cite{zeplin1}.
The details of the experiment are described in Ref. \cite{muons}.
Figure \ref{fig:muonspectrum} shows the spectrum of energy depositions 
compared with the normalised simulated spectrum. The simulations of 
muons were performed taking into account the calculated 
underground muon flux and energy spectrum 
in Boulby rock, the detector geometry, the 
muon energy losses due to ionisation and the propagation of 
scintillation photons in the detector. The calculations of the muon fluxes 
and energy spectra at various depths in Boulby rock were done using
the parameterisation of the muon spectrum at the surface \cite{lvd} and 
the muon propagation code MUSIC \cite{music} (see also Refs. 
\cite{muons,neutrons} for more details).

The simulations were then used to convert the measured muon rate 
in the detector to the muon flux, which was found to be 
$(4.09\pm0.15) \times 10^{-8}$~cm$^{-2}$~s$^{-1}$, where the error includes both 
statistical and systematic uncertainties, the latter being due to the 
applied cuts, uncertainties in the energy threshold, energy 
calibration etc. The comparison between measured and simulated fluxes 
provides an estimate of the rock overburden (column density at 
vertical), which is 2805$\pm$45 m w.~e. assuming a flat surface above 
the detector. The error in depth includes the uncertainties in the 
flux measurements, as well as in the rock composition.

\vspace{-0.3cm}
\section{Simulations of muon-induced neutrons}
\begin{figure}[htb]
\vspace{-0.5cm}
\mbox{
\epsfig{file=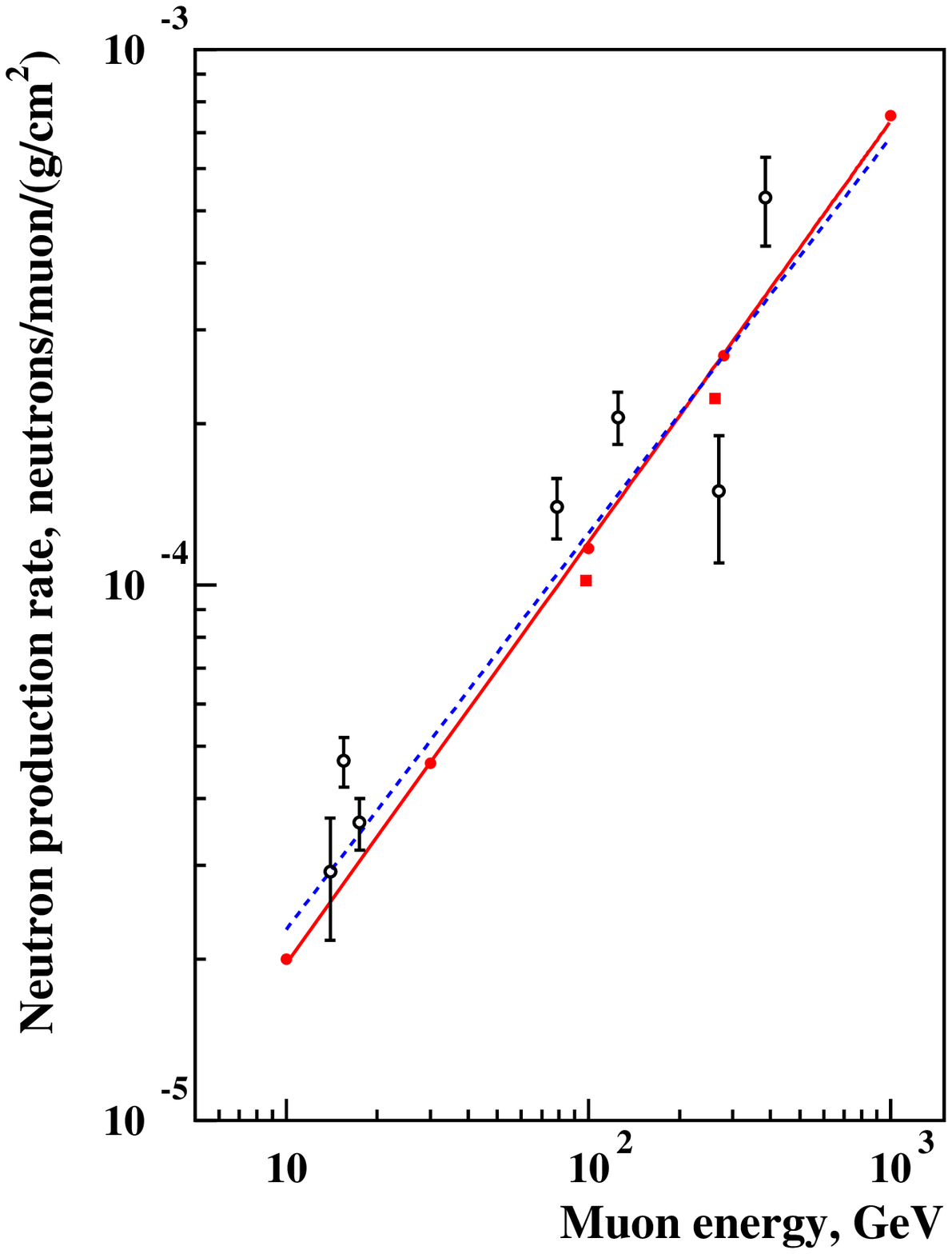,width=5.2cm,height=5.0cm}
\hspace{0.5cm}
\epsfig{file=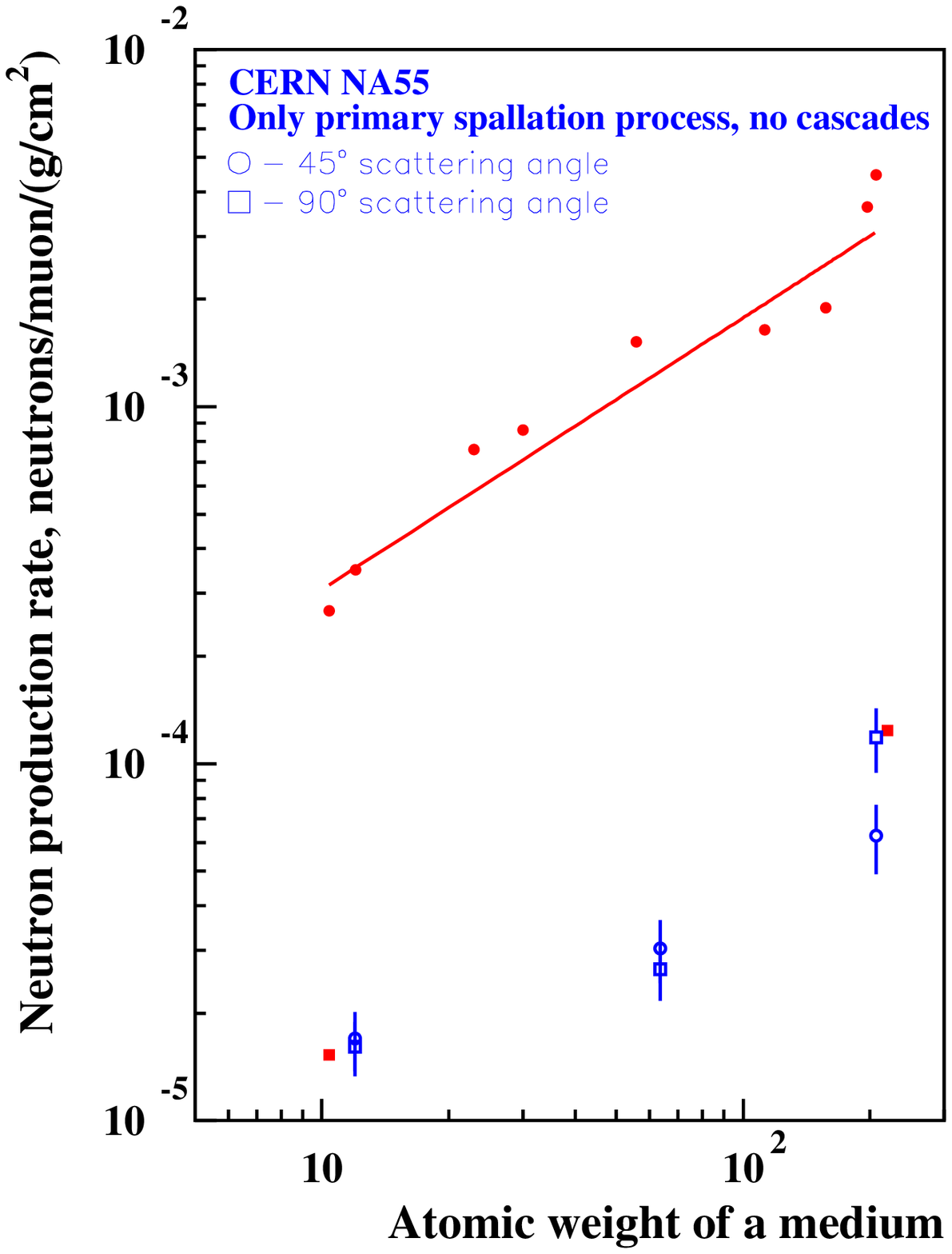,width=5.2cm,height=5.0cm}
}
\parbox{5.2cm}{\vspace{-0.0cm}\caption
{Average number of neutrons produced by a muon per unit path 
length (1 g/cm$^2$) in scintillator as a function of muon energy.
See text for details.
\label{fig:muen}}}
\hspace{0.7cm} \parbox{5.2cm}{\vspace{-0.0cm}\caption
{Dependence of neutron rate on the 
atomic weight of material. For the muon spallation
our result for lead is shifted to
A=220 to avoid the overlaping with the CERN point.
\label{fig:na}}}
\end{figure}

The neutron production by muons and neutron transport were simulated
with FLUKA \cite{fluka}. 
The average number of neutrons produced by a muon per unit path length 
(1 g/cm$^{2}$)
in scintillator is presented in Figure \ref{fig:muen} 
as a function of muon energy. 
Our results (filled circles) have been fitted to a function 
$R_{n} = a \times E^{\alpha}$,
where $a=(3.20 \pm 0.10) \times 10^{-6}$ and $\alpha=0.79 \pm 0.01$.
This is consistent with FLUKA simulations by Wang et al. \cite{wang} (dashed 
line).
Also shown in Figure \ref{fig:muen} are measurements of 
neutron production by 
several experiments (open circles with error bars). 
Their results are plotted as a 
function of mean muon energy for these experiments 
(see Ref. \cite{neutrons} for details).

The simulated neutron production by 
muons with a real spectrum for depths of 0.55 km w.e. and 3 km w.e. 
in Boulby rock (mean energies 98 and 264 GeV, respectively) is shown by
filled squares (below the solid line) in Figure \ref{fig:muen}. In both 
cases a smaller neutron production rate was found. The difference 
is of the order of (10-15)\%. A similar difference was found also 
for muon production in NaCl salt -- the result is 
relevant to the neutron background in salt mines and other rocks.

\vspace{-0.5cm}
We studied also the dependence of neutron rate on the 
atomic weight of material. The neutron rate was obtained with 
280 GeV muons in several materials and compounds and is shown in 
Figure \ref{fig:na} by filled circles. It is obvious that on average 
the neutron rate increases with 
the atomic weight of material, but no exact parameterisation was 
found, which would explain the behaviour for all elements and/or 
compounds. The general trend can be fitted by a simple power-law 
form (solid line in Figure \ref{fig:na}) $R_{n} = b \times A^{\beta}$,
where $b=(5.33 \pm 0.17) \times 10^{-5}$, $A$
is the atomic weight (or mean 
atomic weight in the case of a compound) and $\beta=0.76 \pm 0.01$.
We compared our
results with measurements performed in the NA55 experiment at CERN 
with a 190 GeV muon beam \cite{na}. 
The neutron production was measured in thin 
targets at several neutron scattering angles, so direct comparison 
with our simulations is difficult. The use of thin targets 
allowed the measurement of neutron production in the first muon 
interaction only (without accounting for neutrons produced in cascades)
\cite{na}. We calculated neutron production in the first muon 
interaction in scintillator and lead and plotted it in Figure 
\ref{fig:na} (filled squares) together with the measurements \cite{na} 
at two scattering angles (open circles and open squares). Since the 
measured values refer to particular scattering angles, we normalised 
them to our results at small atomic weight (carbon). The measured 
behaviour of the neutron rate with atomic weight agrees well with 
FLUKA predictions.

\begin{figure}[htb]
\vspace{-0.5cm}
\mbox{
\epsfig{file=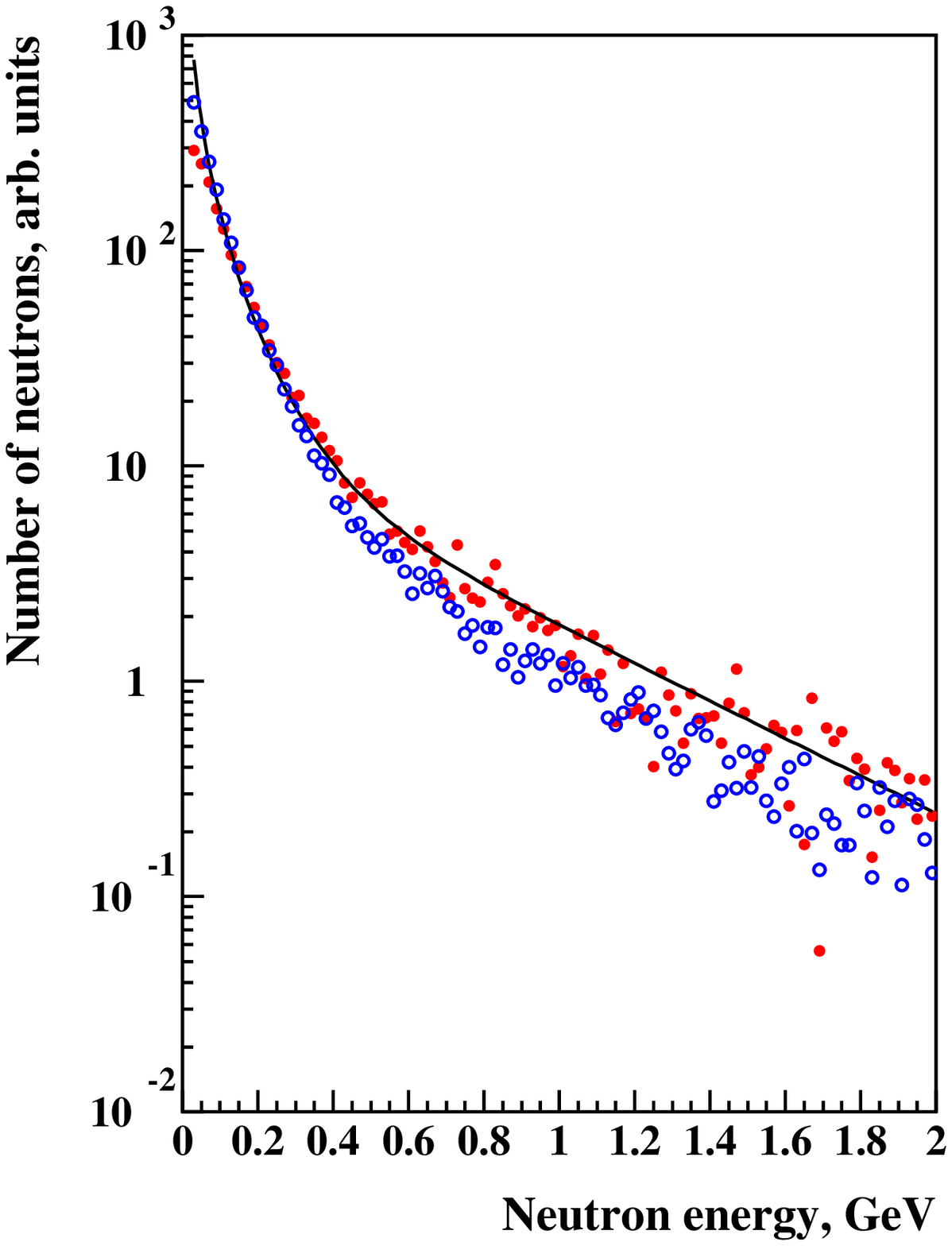,width=5.2cm,height=5.0cm}
\hspace{0.5cm}
\epsfig{file=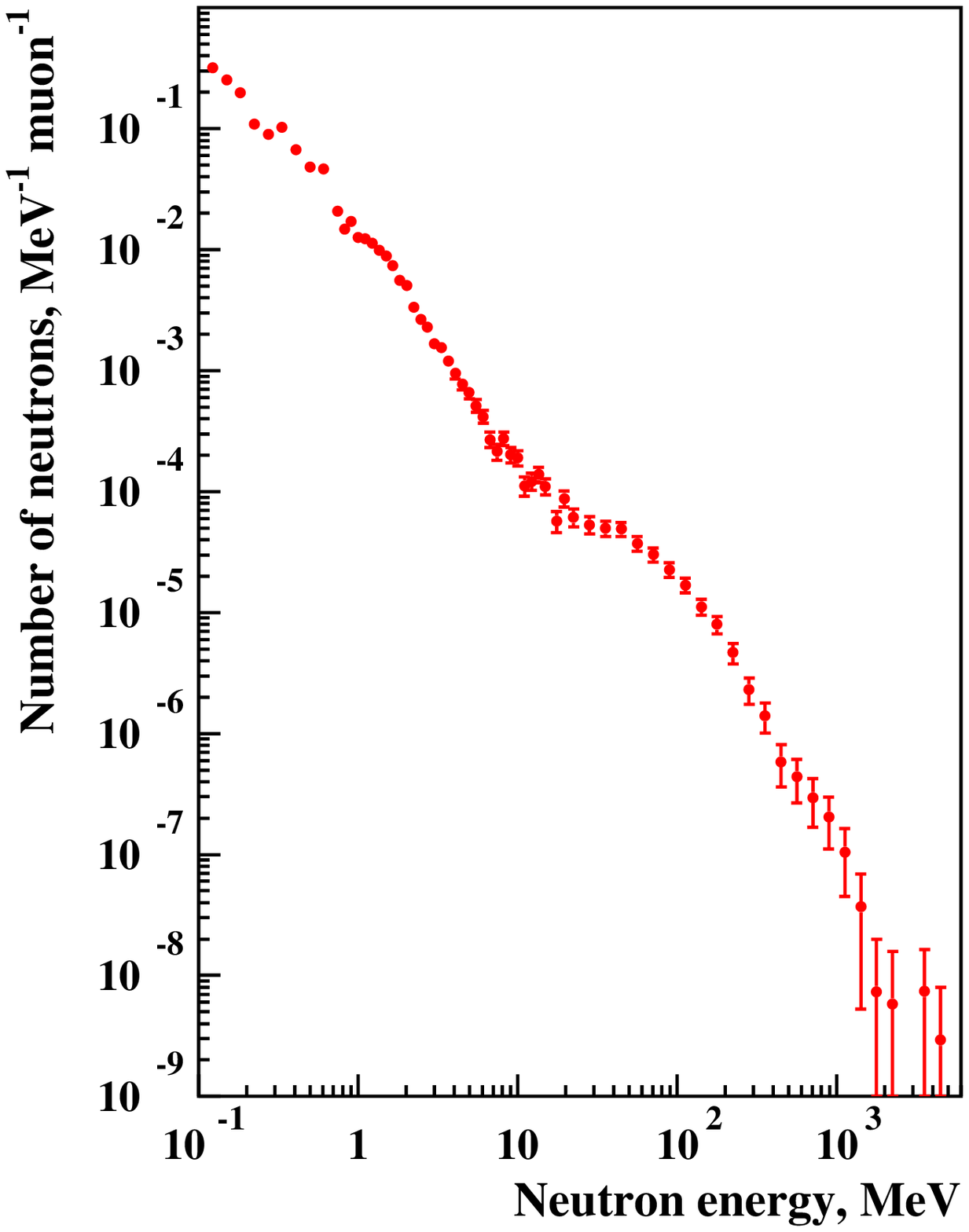,width=5.2cm,height=5.0cm}
}
\parbox{5.2cm}{\vspace{-0.0cm}\caption
{Neutron energy spectrum in scintillator (filled circles) 
and NaCl (open circles).
\label{fig:nspectrum}}}
\hspace{0.7cm} \parbox{5.2cm}{\vspace{-0.0cm}\caption
{Neutron energy spectrum at the boundary between
salt and cavern.
\label{fig:nspectrum2}}}
\end{figure}

\vspace{-0.5cm}
The neutron energy spectrum was calculated for various targets. 
Figure \ref{fig:nspectrum} shows the spectrum obtained for scintillator 
and NaCl together with parameterisation proposed for scintillator and 280 GeV
muons in 
Ref. \cite{wang} (solid curve with arbitrary normalisation). 
In our simulations the real muon spectrum at about 3 km 
w.e. underground was used.  The neutron 
energy spectrum becomes softer with increase of $<A>$, although 
the total neutron production rate increases (see Figure \ref{fig:na}).

The simulation of the energy spectrum of neutrons coming
from the rock (NaCl) into the laboratory hall or cavern was carried out 
with the real muon energy spectrum for Boulby. 
The volume of the salt region was taken as 
$20 \times 20 \times 20$ m$^{3}$, with the cavern 
for the detector of size $6 \times 6 \times 5$ m$^{3}$.
The neutrons in the simulations did not stop in the cavern but were
propagated to the opposite wall where they could be scattered back into
the cavern and could be counted again.
Figure \ref{fig:nspectrum2} shows the simulated neutron 
energy spectrum at the salt/cavern boundary. 
If the neutrons entering the cavern for the first time
are absorbed in the simulations (cannot be scattered back from the
walls and are counted only once), then the flux below 1 MeV
is 2-3 times lower than that shown in Figure \ref{fig:nspectrum2}, in good
agreement with the simulations of neutron propagation by Smith 
\cite{smith}. At high energies the flux does not change much. To obtain the 
neutron flux in units MeV$^{-1}$ cm$^{-2}$ s$^{-1}$ 
the differential spectrum plotted on Figure \ref{fig:nspectrum2}
has to be multiplied by the muon flux.
The total number of 
neutrons entering the cavern is about 
$5.8 \times 10^{-10}$ cm$^{-2}$ s$^{-1}$ above 1 MeV
at 3 km w.~e. in Boulby rock. 
The flux on an actual detector can
be different from the flux on the boundary salt/cavern due to the
interactions of neutrons in the detector itself.

\vspace{-0.3cm}
\section{Conclusions}
We have discussed neutron background, in particular muon-induced 
neutron background relevant to dark 
matter experiments.
Neutron production by cosmic-ray muons was simulated for various 
muon energies and various materials. We found reasonably good 
agreement with the recent experimental data. 
Our simulation is the starting point of a 
three-dimensional Monte Carlo to study the neutron
background for any detector.

\vspace{-0.3cm}
\section*{Acknowledgments}
We wish to thank PPARC for financial support.
We are also grateful to the UKDMC and to the staff of Cleveland Potash Ltd
for technical assistance.
P. F. Smith acknowledges support from an Emeritus Research Award by the 
Leverhulme Trust.

\end{document}